\documentclass[fleqn,10pt]{wlscirep}
\usepackage[T1]{fontenc}
\usepackage[utf8]{inputenc} 
\usepackage{calc}
\usepackage{textcomp}
\usepackage{graphicx}
\usepackage[caption=false]{subfig}
\usepackage[none]{hyphenat}
\usepackage{bm}
\usepackage{epsfig}
\usepackage{hyperref}
\usepackage{dsfont}
\usepackage{braket}

\title{Experimental quantum state discrimination using the optimal fixed rate of inconclusive outcomes strategy}
\author[1,2]{Santiago G\'omez}
\author[1,2,*]{Esteban S. G\'omez}
\author[3,+]{Omar Jim\'enez}
\author[1,2]{Aldo Delgado}
\author[1,2]{Stephen P. Walborn}
\author[1,2]{Gustavo Lima}

\affil[1]{Departamento de F\'isica, Universidad de Concepci\'on, casilla 160-C, Concepci\'on, Chile}
\affil[2]{Millennium Institute for Research in Optics, Universidad de Concepci\'on, casilla 160-C, Concepci\'on, Chile}
\affil[3]{Centro de \'Optica e Informaci\'on Cu\'antica, Facultad de Ciencias, Universidad Mayor, Camino La Pirámide N$^{\circ}$5750, Huechuraba, Santiago, Chile}
\affil[*]{estesepulveda@udec.cl}
\affil[+]{omar.jimenez@umayor.cl}
\date{\today}

\begin{abstract}
The problem of non-orthogonal state discrimination underlies crucial quantum information tasks, such as cryptography and computing protocols. Therefore, it is decisive to find optimal scenarios for discrimination among quantum states. We experimentally investigate the strategy for the optimal discrimination of two non-orthogonal states considering a fixed rate of inconclusive outcomes (FRIO). The main advantage of the FRIO strategy is to interpolate between unambiguous and minimum error discrimination by solely adjusting the rate of inconclusive outcomes. We present a versatile experimental scheme that performs the optimal FRIO measurement for any pair of generated non-orthogonal states with arbitrary a priori probabilities and for any fixed rate of inconclusive outcomes. Considering different values of the free parameters in the FRIO protocol, we implement it upon qubit states encoded in the polarization mode of single photons generated in the spontaneous parametric down-conversion process. Moreover, we resort to a newfangled double-path Sagnac interferometer to perform a three-outcome non-projective measurement required for the discrimination task, showing excellent agreement with the theoretical prediction. This experiment provides a practical toolbox for a wide range of quantum state discrimination strategies using the FRIO scheme, which can greatly benefit quantum information applications and fundamental studies in quantum theory.
\end{abstract}

\begin{document}
\flushbottom
\maketitle
\thispagestyle{empty}
\noindent Quantum measurements lie at the core of quantum mechanics, and are a cornerstone of  interpretations of quantum theory \cite{peres_quantum_1993}. Moreover, they have a crucial role in the evolution of quantum systems \cite{kraus_states_1983} and have found appealing applications such as estimating unknown physical parameters using quantum resources \cite{T_th_2014,Fabio2020}. In quantum information science, measurements are especially relevant for implementing quantum computing and communication protocols \cite{Nielsen}. The simplest scenario involves two parties, where a sender can prepare and send  information encoded in quantum states \cite{Bae_2015}. To access to this information, the receiver has to choose which quantum measurement will be performed \cite{Bergou,Barnett2}. Usually, the chosen measurement will depend on the properties of the received states and the features of the quantum protocol to be implemented. One possible task is identifying an unknown quantum state. One way to address this situation lies in the quantum tomography technique, which reconstructs the quantum state of an unknown physical system from the measured probabilities of a suitable set of observables. Nevertheless, the number of measurements required for a successful state reconstruction scales at least polynomially in the dimension of the state \cite{Kwiat2001}, and requires multiple identical copies. Other tasks require identifying quantum states among others belonging to a given set in a single-shot measurement. However, performing this assignment is impossible deterministically when non-orthogonal states are considered. 

Hence, quantum state discrimination (QSD) relies on identifying a quantum state belonging to a set of $N$ known non-orthogonal quantum states. This task plays a fundamental role in several remarkable quantum protocols such as quantum key distribution \cite{BennettQKD}, quantum teleportation \cite{Ivette,Neves2}, entanglement swapping \cite{Miguel-Angel,PhysRevA.71.012303}, and entanglement concentration \cite{Chefles,PhysRevA.87.052327}. Moreover, being a fundamental protocol, QSD also has been studied in relation with: contextuality \cite{Spekkens}, path distinguishability \cite{Bera,Bagan,njp.11.073035}, and quantum correlations \cite{OmarJ,Khalid,e23010073}. Thus, there are well-known strategies to implement QSD, namely the \emph{minimum error discrimination} (MED) \cite{Holevo,Helstrom}, the \emph{unambiguous discrimination} (UD) \cite{Ivanovic,Dieks,Peres3,Jimenez2}, and the \emph{maximum confidence discrimination} (MCD) \cite{CrokeMC,JimenezMC,HerzogMC}, each focused on optimizing some figure of merit \cite{AChefles}. To implement any QSD strategy, it is necessary first to determine the corresponding measurement defined by a set of positive operator-valued measures (POVMs) \cite{Chefles-Barnett,He}, and then engineer a way to implement these POVMs in an experimental realization \cite{Reck,BergouE}. In the case of MED, there are $N$ POVM elements $\Pi_i$ associated with the discrimination of one of the states $\rho_i$ \cite{Bergou}. However, since the given states are non-orthogonal, the discrimination process will inevitably introduce some error in the identification of each quantum state $\rho_i$ \cite{Holevo}. In the MED strategy, the average error probability is minimized \cite{Helstrom,Han}. Conversely, for UD, there is no error in identifying each quantum state $\rho_i$. This process can be realized by introducing an additional POVM element, $\Pi_0$, which is associated with an inconclusive outcome \cite{Peres3}. Finally, the MCD strategy maximizes the confidence in taking the measurement outcome $i$ to indicate that the $\rho_i$ state was prepared \cite{Barnett2}. QSD has been experimentally implemented for MED \cite{Clarke, Barnett,Higgins,Lu,Waldherr,Neves}, UD \cite{Clarke2,Mohseni,Miguel-Angel_2021,PhysRevA.94.042309} and also for MCD \cite{Mosley2,Steudle}. Moreover, the experimental realization for maximizing the mutual information \cite{Fields} between two users was also performed \cite{Mizuno}. 

Under certain conditions, the MED and UD strategies coincide with the MCD strategy \cite{Gina,JimenezMC}. Moreover, MED and UD can be joined simultaneously in a more general QSD scheme known as fixed rate of inconclusive outcomes (FRIO) \cite{Chuan,Eldar,Jaromir}. In the case of QSD by FRIO, the average error probability in the identification of the quantum state $\rho_i$ is minimized under the condition of fixing the probability of inconclusive results \cite{Ulrike2}. Although for MED, UD and MCD there is no analytical solution to the discrimination by FRIO of $N$ arbitrary non-orthogonal states \cite{Ulrike}, in the case of two pure non-orthogonal states with arbitrary a \textit{priori} preparation probabilities, the complete optimal solution is known \cite{Gina, JimenezFRIO}. Moreover, experimental schemes for the realization of FRIO onto two pure non-orthogonal states have been proposed \cite{Hai,Shehu}.

Here we present an experimental realization of FRIO discrimination on two pure non-orthogonal states encoded on the polarization state of single photons created by the spontaneous parametric down-conversion (SPDC) process. We use a double-path Sagnac interferometer to implement a genuine three-outcome POVM over a single polarization qubit performing the required measurement. A notable feature of our experimental setup is that it can be easily configured to implement the optimal FRIO POVM for all pairs of non-orthogonal states, arbitrary a priori probabilities, and any value of the fixed rate of inconclusive outcome. In particular, it allows us to implement MED, UD, and any intermediate case of FRIO by fixing the rate of inconclusive outcomes, showing a good agreement with the theoretical results \cite{Gina,JimenezFRIO}. This experimental platform gives a flexible and customizable toolbox for a wide range of QSD strategies involving the FRIO scheme, which can be helpful to explore novel applications for quantum information protocols and further research in quantum foundations. 

\section*{Methods}   \label{sec:frio}
We consider a qubit source of two pure non-orthogonal states denoted by $|\phi_1\rangle$ and $|\phi_2\rangle$, with \textit{a priori} preparation probabilities $\eta_1$ and $\eta_2$, respectively.  Without loss of generality, these states can be written as 
\begin{align*}
|\phi_1\rangle&=\cos\alpha|0\rangle+\sin\alpha|1\rangle,\\
|\phi_2\rangle&=\cos\alpha|0\rangle-\sin\alpha|1\rangle,
\end{align*}
where the states $\{|0\rangle, |1\rangle\}$ represent the logical basis. The overlap between the states is given by the parameter $s=\langle\phi_1|\phi_2\rangle=\cos(2\alpha)$, considering $s\in[0,1]$. Moreover, the \textit{a priori} probabilities can be set arbitrarily and they must satisfy the constraint $\eta_1+\eta_2=1$. The FRIO discrimination process is carried out by using three POVM elements, where $\Pi_{1(2)}$ is associated to the identification of $|\phi_{1(2)}\rangle$ and $\Pi_0$ corresponds to the inconclusive outcome, where no information of the states can be learned from the measurement. These three operators satisfy the condition
\begin{equation*}\label{eq:povm}
\Pi_{1}+\Pi_{2}+\Pi_{0}=\mathds{1},
\end{equation*}
where $\mathds{1}$ is the identity operator. We define the following probabilities $p_{1(2)}$, $r_{1(2)}$ and $q_{1(2)}$, corresponding to the probabilities of success, error and the inconclusive outcome, respectively, in the discrimination of $|\phi_{1(2)}\rangle$. These probabilities encompass all possible outcomes in this case, that is $p_{1(2)}+r_{1(2)}+q_{1(2)}=1$.

The average probabilities of success $P_s$, error $P_e$ and inconclusive outcome $Q$ over the states, are given by \cite{Gina}
\begin{equation}
\begin{aligned}\label{ecu:Probb}
P_s&=tr(\eta_1\rho_1\Pi_1)+tr(\eta_2\rho_2\Pi_2)=\eta_1p_{1}+\eta_2p_{2},\\
P_e&=tr(\eta_1\rho_1\Pi_2)+tr(\eta_2\rho_2\Pi_1)=\eta_1r_1+\eta_2r_2,\\
Q&=tr(\rho\Pi_0)=\eta_1q_1+\eta_2q_2,
\end{aligned}
\end{equation}
where $\rho_{1(2)}=|\phi_{1(2)}\rangle\langle\phi_{1(2)}|$, and $\rho=\eta_1\rho_1+\eta_2\rho_2$. It is straightforward to see that the average probabilities satisfy $P_s+P_e+Q=1$.

\subsection*{Optimal strategy for FRIO discrimination}
The optimal FRIO strategy minimizes the average error probability $P_e$ under the constraint that the inconclusive outcome probability $Q$ is fixed. For this case, the optimal probabilities for FRIO were obtained by Bagan \textit{et al.}\cite{Gina}. Due to the symmetry of this task, it is enough to consider the case $\eta_1\leq \eta_2$, that is, when $0\leq \eta_1\leq 1/2$. The FRIO solution identifies three intervals where the optimal probabilities can be obtained. These intervals depend on the values of the overlap $s$ and the probability $\eta_1$ given a fixed value of $Q$ \cite{Gina,JimenezFRIO}. Interval I is defined as
\begin{eqnarray*}
\frac{s^2}{1+s^2}\leq \eta_1\leq 1/2\hspace{0.5cm}\text{and}\hspace{0.5cm}
0\leq Q\leq Q_0,
\end{eqnarray*} 
where $Q_0=2s\sqrt{\eta_1\eta_2}$ is the maximum inconclusive event probability for this interval. 
Interval II is defined as
\begin{eqnarray*}
0\leq \eta_1\leq \frac{s^2}{1+s^2}\hspace{0.5cm}\text{and}\hspace{0.5cm}
0\leq Q\leq Q_{th},
\end{eqnarray*}
with $Q_{th}=\frac{2\eta_1\eta_2(1-s^2)}{1-Q_0}$  a threshold value which separates intervals II and III. The optimal probabilities for these intervals are given by
\begin{align*}
q_i&=\frac{Q}{2\eta_i},\\
r_i&=\frac{1}{2}\left(1-q_i-\frac{(1-q_i)\overline{Q}-\frac{(Q_0-Q)^2}{2\eta_i}}{\sqrt{\overline{Q}^2-(Q_0-Q)^2}}\right),\\ 
p_i&=1-q_i-r_i,
\end{align*}
for $i=1,2,$ and $\overline{Q}=1-Q$. Therefore, the optimal error probability $P_e$ valid in the intervals I and II is minimal, and reads
\begin{equation*}
P_e=\frac{1}{2}\left(\overline{Q}-\sqrt{\overline{Q}^2-(Q_0-Q)^2}\right).
\end{equation*}

On the other hand, the interval III is defined when
\begin{eqnarray*}
0\leq \eta_1\leq \frac{s^2}{1+s^2}\hspace{0.5cm}\text{and}\hspace{0.5cm} Q_{th}\leq Q\leq \eta_1+\eta_2s^2,
\end{eqnarray*} 
and the optimal probabilities are given by
\begin{eqnarray*}
p_1&=&0,\hspace{0.3cm}
r_1=\frac{P_e}{\eta_1},\hspace{0.3cm}
q_1=1-r_1,\hspace{0.3cm}r_2=0,\\
p_2&=&\left(s\sqrt{r_1}+\sqrt{(1-r_1)(1-s^2)}\right)^2,\hspace{0.15cm} q_2=1-p_2,
\end{eqnarray*}
and the optimal average error probability is given by \cite{JimenezFRIO}
\begin{equation*}
P_e=\frac{1}{1-4c}\left(\eta_1\overline{Q}+c(\eta_2-\eta_1-2\overline{Q})-Q_0\sqrt{c(Q\overline{Q}-c)}\right),
\end{equation*}
where $c=\eta_1\eta_2(1-s^2)$. Note that the optimal strategy in the interval III is implemented using a two-outcome projective measurement. 

A notable feature of FRIO discrimination is that allow not only MED (when $Q=0$) and UD (when $Q=Q_{max}$), but also intermediate cases. Here, $Q_{max}$ is the maximum possible value for $Q$ in each interval defined above. To show the versatility of this state discrimination scheme, we present an optical experiment performing FRIO for both cases (MED and UD), and we consider an intermediate case for the inconclusive event $Q=Q_0/2$ using the same device implementing the required POVMs in Eq. (\ref{ecu:Probb}) for each interval. The optimal POVM implementation depends on the overlap $s$ between the non-orthogonal qubit states $\{|\phi_i\rangle\}$. These states are prepared encoding on the polarization modes (such as horizontal and vertical directions) of single photons generated using an heralded source based on the SPDC process. We are able to finely tuning the parameter $s$ and the a \emph{priori} preparation probabilities $\eta_i$, performing FRIO strategies ($Q=0$, $Q=Q_0/2$, and $Q=Q_{max}$) considering seven values of the overlap $s$ in two preparation scenarios, when $\eta_1=\eta_2$ and $\eta_1<\eta_2$.

\subsection*{Experimental description}\label{sec:experimetal}
\begin{figure*}[!t]
\centering
 \includegraphics[width=\textwidth]{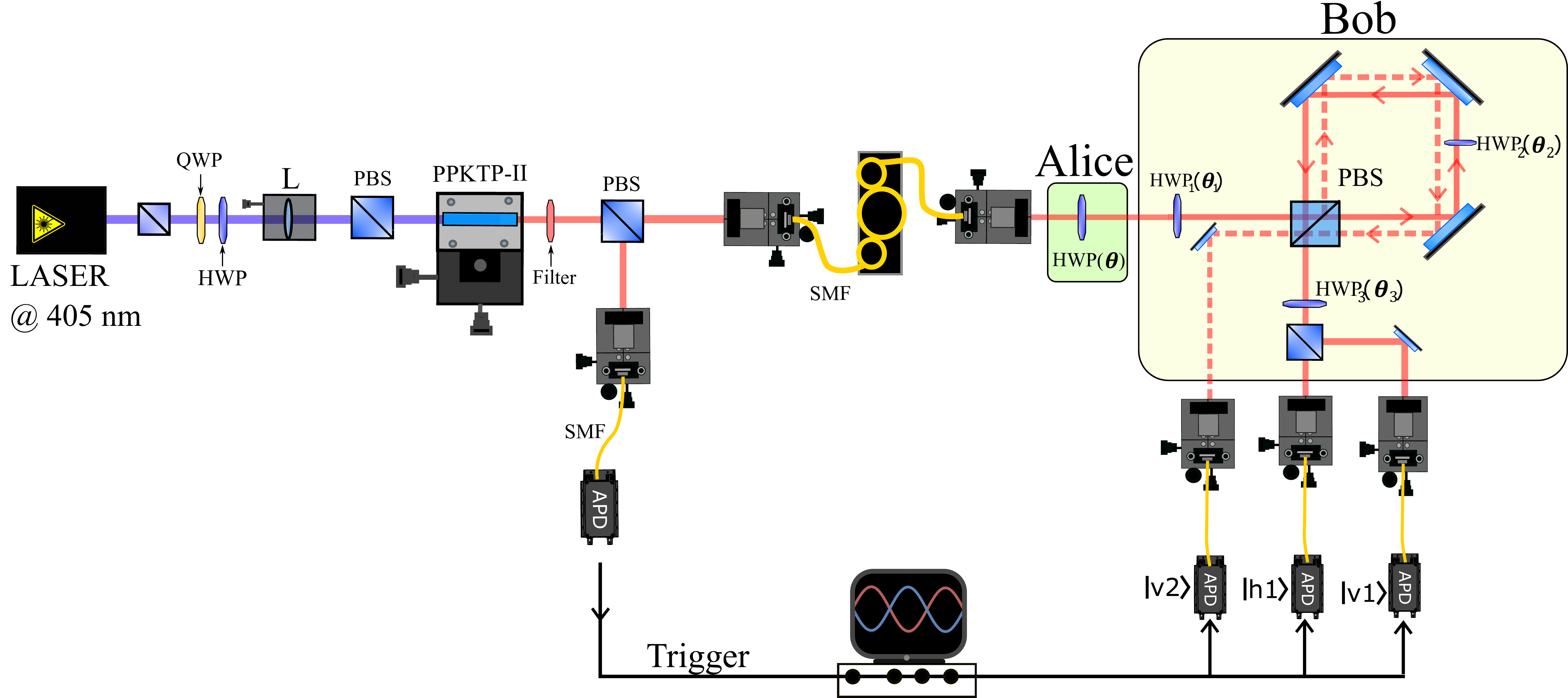}
 \caption{Experimental setup used to implement FRIO quantum state discrimination. Alice can prepares either the non-orthogonal state $\ket{\phi_i}$ using the half-wave plate $HWP(\theta)$, encoding on a single photon generated at the heralded source based on the SPDC process. Bob uses a two-path Sagnac interferometer to perform the three-outcomes POVM needed for the chosen discrimination strategy covered by the FRIO scheme. Each outcome is associated with the success, error, and inconclusive result. In particular, for intervals I and II, the inconclusive outcome $Q$ are detected in the spatial mode labeled by $\ket{V2}$. Thus, the state $\ket{\phi_{1}}$ is identified when $\ket{V1}$ is detected, while the $\ket{\phi_{2}}$ is associated to detecting in $\ket{H1}$. See the main text for more details.}\label{setup}
\end{figure*}

To carry out FRIO quantum state discrimination, we implement a heralded photon source based on the spontaneous parametric down conversion (SPDC) process and a two-path Sagnac interferometer to perform the required measurement. The experimental setup is depicted in Fig. \ref{setup}. A continuous-wave laser at $405$ nm pumps a type-II nonlinear periodically poled potassium titanyl phosphate (PPKTP) crystal to create degenerate down-converted photons at $810$ nm with horizontal and vertical polarization. To ensure the degenerate phase-matching conditions and remove the remaining pump beam, a Semrock high-quality narrow bandpass filter centered at $810$ nm is used, with $0.5$ nm of bandwidth and a peak transmission of $> 90\%$. To maximize the coincidence count rate, we consider a numerical model\cite{LT05}. Precisely, the optimal coupling condition is reached when $\omega_{DCP}=\sqrt{2}\omega_{p}$, where $\omega_{p}$ and $\omega_{DCP}$ are the waist modes of the pump beam and the down-converted photons at the center of the PPKTP crystal, respectively. In our case, these waists are adjusted by using a $20$ cm focal length lens $L$ and $10\times$ objective lenses. 
The generated photon pairs can be used to implement a heralded single-photon source, in the sense that one down-converted photon arrives at the trigger detector announcing the passage of the other photon through the stages of Alice and Bob (see Fig. \ref{setup}). Therefore, a polarizing beam splitter (PBS) is placed after the PPKTP crystal to separate the down-converted photons deterministically. Then, they are sent to the trigger and Alice section coupling into single-mode optical fibers (SMF), removing any spatial correlation between them that could arise from imperfections when satisfying the phase-matching conditions in the crystal. Moreover, to maintain the polarization state of the photons through the propagation in the optical fiber, we use a manual fiber polarization controller for the photon arriving at Alice, and polarizing films are placed in front of the trigger and Alice's detectors to ensure the correct polarization mode of the detected state.

Alice can prepare the two non-orthogonal states $\{\ket{\phi_1},\ket{\phi_2}\}$ using the half-wave plate HWP($\theta$) to encode a polarization qubit her photon. The polarization qubit states read
\begin{equation}
\begin{aligned}
|\phi_1\rangle&=\cos2\theta|H\rangle+\sin2\theta|V\rangle,\\
|\phi_2\rangle&=\cos2\theta|H\rangle-\sin2\theta|V\rangle,\label{qubits}
\end{aligned}
\end{equation}
where $|H\rangle$ and $|V\rangle$ are the horizontal and vertical polarization modes, and $\theta$ is the inclination angle of the HWP with respect to its fast axis. The photons are sent to Bob through free space to implement the FRIO state discrimination procedure. To generate the global unitary transformation $U$ over the non-orthogonal states requierd to implement the POVM \cite{Shehu}, we resort to couple the polarization degree of freedom with two spatial propagation modes as an ancilla system. First, Bob rotates the polarization state using a half-wave plate HWP$_{1}$ oriented at angle $\theta_{1}$ and then inputs the state to the two-path Sagnac interferometer configuration, which is composed of three laser mirrors, the HWP$_{2}$ and a PBS. In this device the polarization is coherently coupled with the spatial modes \cite{almeida07,GMGCFAL18, GGGCBD16}, since the PBS splits the incident photon through the clockwise (reflected) or counterclockwise (transmitted) mode inside the interferometer. Thus, the PBS operation can be seen as a controlled-NOT gate: the photon populates a spatial mode depending on the input polarization state. Moreover, in the counterclockwise path there is HWP$_{2}$ aligned at angle $\theta_{2}$, which rotates again the photon's polarization if it propagates in this spatial mode. Then, a new passage through the same PBS superposes the two spatial modes. We denote the initial and second photon propagation paths by $|1\rangle$ and $|2\rangle$. Lastly, a HWP$_{3}$ at $\theta_{3}$ and a PBS are placed in the $|1\rangle$ mode to obtain three outcomes which finally determine the required POVM. The global unitary can be written as
\begin{equation*}
U=C(\theta_{3}) \cdot CNOT \cdot C(\theta_{2}) \cdot CNOT \cdot C(\theta_{1}),
\end{equation*}
where the transformation $C(\theta_{i})$ represents a rotation of the polarization in an angle $2\theta_{i}$. Then, applying $U$ onto the non-orthogonal states $\{|\phi_1\rangle,|\phi_2\rangle\}$ gives us the following transformation,
\begin{equation}
\begin{aligned}
U|\phi_1\rangle|1\rangle&=\sqrt{p_1}|V\rangle|1\rangle+\sqrt{r_1}|H\rangle|1\rangle+\sqrt{q_1}|V\rangle|2\rangle,\\
U|\phi_2\rangle|1\rangle&=\sqrt{r_2}|V\rangle|1\rangle+\sqrt{p_2}|H\rangle|1\rangle+\sqrt{q_2}|V\rangle|2\rangle,\label{Us}
\end{aligned}
\end{equation}
where the parameters $p_i$, $r_i$, and $q_i$ are the optimal success, error, and inconclusive outcome probabilities associated to discriminate $|\phi_i\rangle$. Indeed, for the case when $\eta_{1}=\eta_{2}=1/2$ (that is, equal state preparation), we can write these optimal probabilities in terms of the waveplate angles $\theta$ and $\theta_1$, $\theta_2$, $\theta_3$:
\begin{equation}
\begin{aligned}\label{eq:probI}
p_i& = \frac{1}{2}(\cos2\theta\cos\theta_{2}-\sin2\theta)^{2},\\
r_i& = \frac{1}{2}(\cos2\theta\cos\theta_{2}+\sin2\theta)^{2},\\
q_i& = (\sin\theta_{2}\cos2\theta)^{2},
\end{aligned}\end{equation}
for $i=1,2$, while $\theta_{1}$ and $\theta_{3}$ are fixed at $0$ and $\pi/4$, respectively. Remarkably, for these values of the initial state preparation probabilities $\eta$, the optimal probabilities for FRIO discrimination correspond to the interval I for any inner product $s$ between the states $|\phi_i\rangle$.

On the other hand, for $\eta_1<\eta_2$ we have that the optimal probabilities are given by
\begin{equation}
\begin{aligned}\label{eq:probIyII}
p_{1},r_{2}& = (\cos2\theta(\sin\theta_{3}\cos\theta_{2}\cos\theta_{1}-\cos\theta_{3}\sin\theta_{1})\pm\sin2\theta(\sin\theta_{3}\cos\theta_{2}\sin\theta_{1}+\cos\theta_{3}\cos\theta_{1}))^{2},\\
p_{2},r_{1}& = (\cos2\theta(\cos\theta_{3}\cos\theta_{2}\cos\theta_{1}+\sin\theta_{3}\sin\theta_{1})\pm\sin2\theta(\sin\theta_{3}\cos\theta_{1}-\cos\theta_{3}\cos\theta_{2}\sin\theta_{1}))^{2},\\
q_{1,2}& = \sin\theta_{2}(\cos2\theta\cos\theta_{1}\pm\sin2\theta\sin\theta_{1})^{2},
\end{aligned}
\end{equation}
where the sign $\pm$ is taking according the state labeled by $i=1,2$ to be discriminated in the FRIO process. Hence, the action of $U$ reveals the three possible results associated to the output states which we labeled as $\{|V\rangle|1\rangle,\,|H\rangle|1\rangle,\,|V\rangle|2\rangle\}$. Thus, detection at the corresponding output modes is the final step for implementing any POVM described by the operation given in Eq. (\ref{Us}). For FRIO discrimination, the unitary transformation in Eq. (\ref{Us}) allows us to cover all cases belonging to the intervals I and II discussed in the last section. Indeed, detection in the $|V\rangle|2\rangle$ mode corresponds to the inconclusive result for both states $|\phi_i\rangle$ (see Fig.\ref{setup}). On the other hand, detection in the $|V\rangle|1\rangle$ mode corresponds to the success (error) in the discrimination of $|\phi_{1(2)}\rangle$, while a detection in $|H\rangle|1\rangle$ corresponds to the error (success) in the discrimination of $|\phi_{1(2)}\rangle$. 

For the case of interval III, the optimal measurement corresponds to a two-outcome projection \cite{Gina}. In this case, the unitary transformation can be written as follows
\begin{align}
U|\phi_1\rangle|1\rangle&=\sqrt{q_1}|V\rangle|1\rangle+\sqrt{r_1}|H\rangle|1\rangle,\nonumber\\
U|\phi_2\rangle|1\rangle&=\sqrt{q_2}|V\rangle|1\rangle+\sqrt{p_2}|H\rangle|1\rangle,\label{Ust}
\end{align}
where there is no detection at $|V\rangle|2\rangle$ mode. Thus, the detection in $|V\rangle|1\rangle$ corresponds now with the inconclusive result. Then, for $\eta_1<\eta_2$ in the interval III we obtain the following optimal probabilities:
\begin{equation}
\begin{aligned}\label{eq:probIII}
p_{2},r_{1}& = (\cos2\theta\cos\theta_{3}\pm\sin2\theta\sin\theta_{3})^{2},\\
q_{1}& = (\cos2\theta\sin\theta_{3}\pm\sin2\theta\cos\theta_{3})^{2}.\\
\end{aligned}
\end{equation}

To detect the output photons, PerkinElmer single-photon avalanche detectors (APDs) were placed in the trigger path and the outputs of the Sagnac interferometer to record the photon statistics. A coincidence count module receives the signal from the detectors, where the timing delay was adjusted between each detector's output and the heralding trigger signal. We actively control the pump laser power (1 mW), setting a 500 ps coincidence gate to minimize the accidental counts, generating a coincidence rate of $\sim 1400$ photons pairs per second. This corresponds to a spectral brightness up to $\sim 400000$ photon pairs (s mW nm)$^{-1}$.

From the above discussion, it is clear that our experimental setup allows us to implement the optimal FRIO POVM for any pair of non-orthogonal states, arbitrary a priori probabilities, and any value of the fixed rate of inconclusive outcomes. This is done by adjusting the values of the angles $\theta_1, \theta_2$ and $\theta_3$ on the corresponding wave plates.

\section*{Results and Discussion}\label{sec:Results}
\begin{figure*}[!t]
\centering
 \includegraphics[width=1\textwidth]{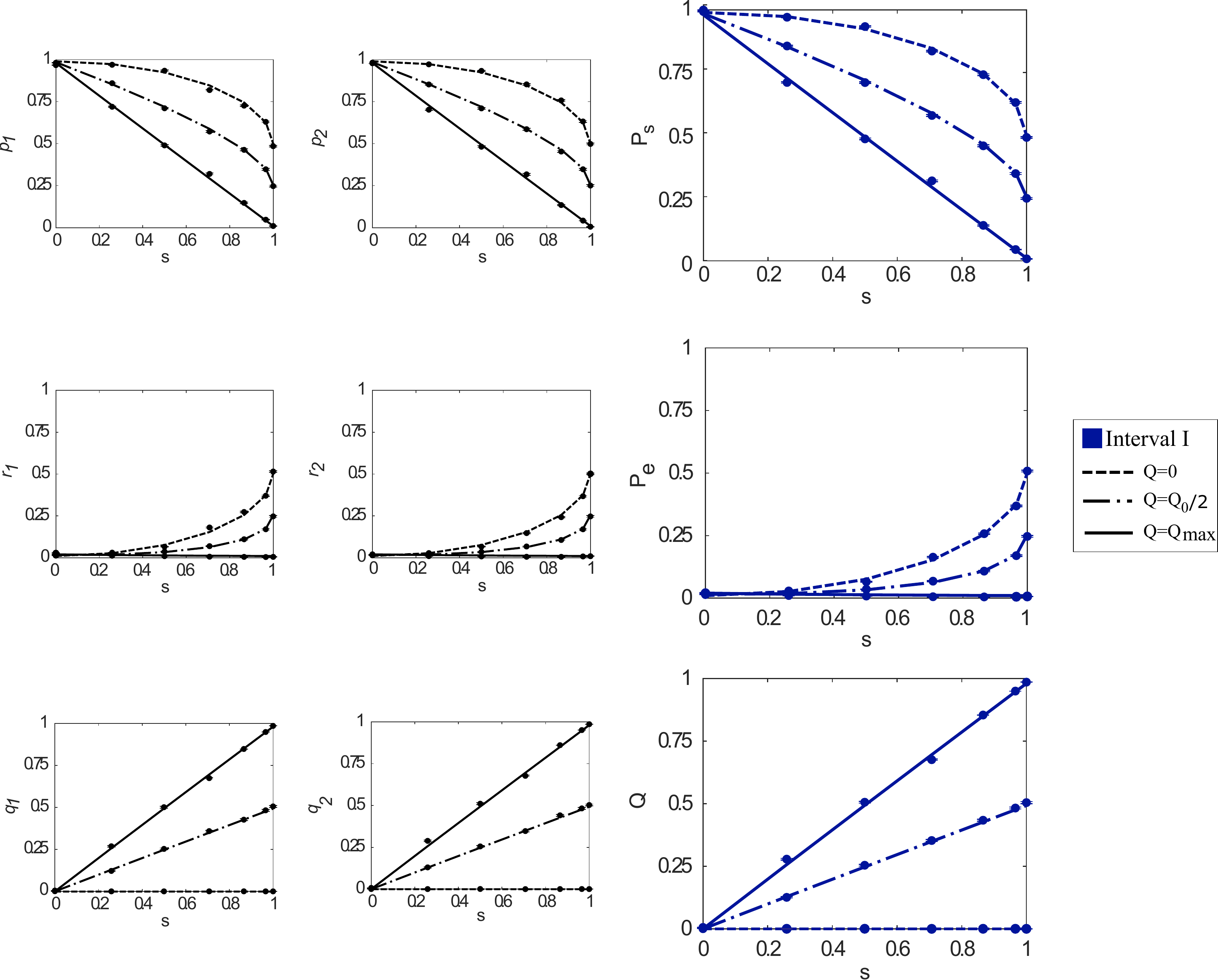} \caption{The success $P_{s}$, error $P_{e}$ and inconclusive $Q$ probables as a function of inner product $s$ when $\eta_{1}=\eta_{2}=0.5$ for both states $\ket{\phi_i}$. The lines in each subplot represent the considered values for $Q$ and were computed considering the state in Eq. (\ref{rho}).}
 \label{iqual}
\end{figure*}

\begin{figure*}[!t]
 \includegraphics[width=1\textwidth]{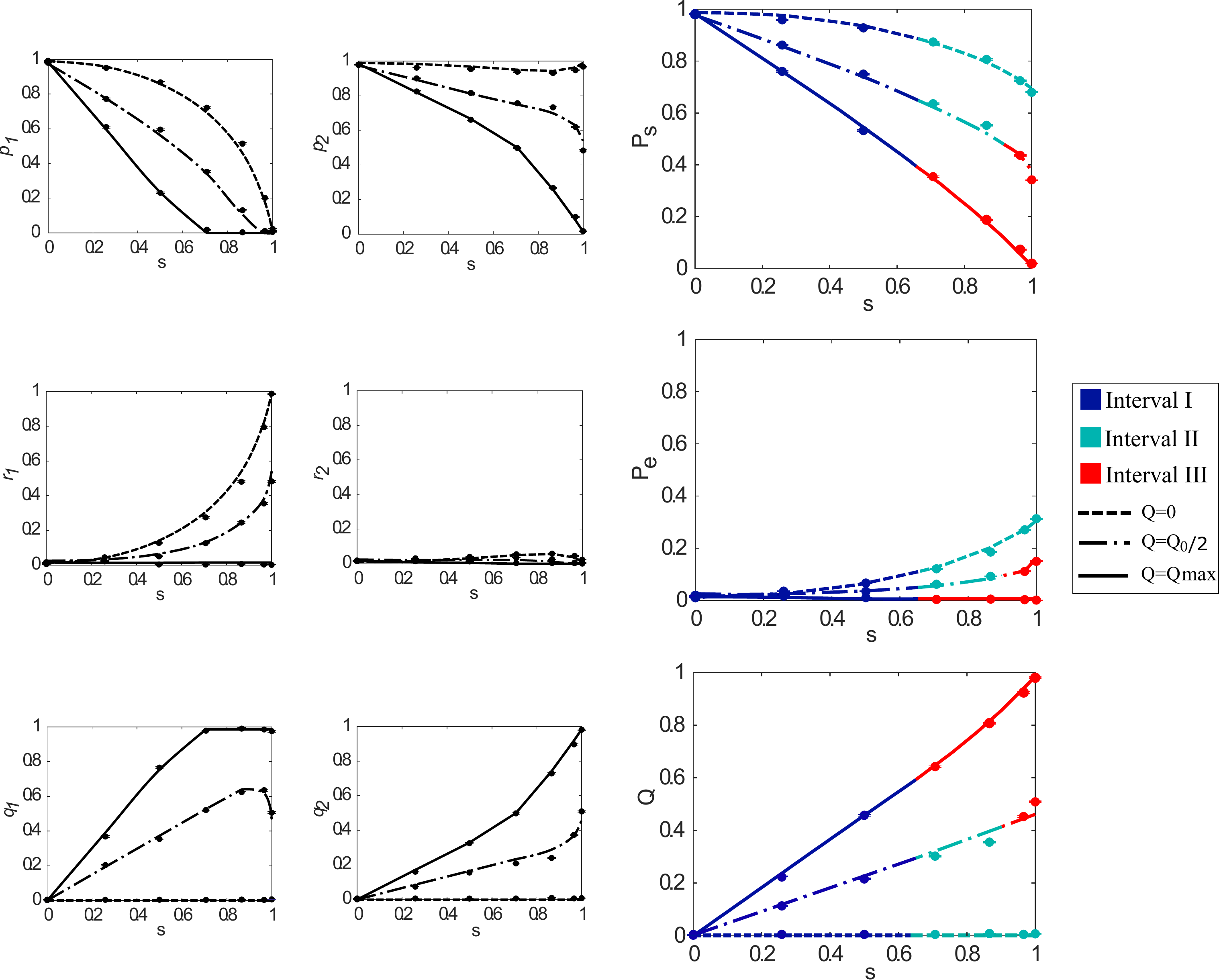} \caption{The success $P_{s}$, error $P_{e}$ and inconclusive $Q$ probables as a function of inner product $s$ when $\eta_{1}=0.3$ and $\eta_{2}=0.7$ for $\ket{\phi_{1}}$ and $\ket{\phi_{2}}$ respectively. The considered values for $Q$ are depicted by different lines in each subplot and were obtained from the state given in Eq. (\ref{rho}).}
 \label{diffe}
\end{figure*}

To show the practicality of the FRIO discrimination scheme using our experimental setup, we implemented the discrimination procedure between the two non-orthogonal polarization qubits given in Eq. (\ref{qubits}) for a range of state overlaps $s$. A key feature of our experiment is that we use a single optical device to implement the complete range of FRIO from MED to UD and including any intermediate case. Indeed, the implementation of every strategy depends on the Sagnac interferometer configuration relying on the angles of the half-wave plates HWP$_{1}$($\theta_{1}$), HWP$_{2}$($\theta_{2}$) and HWP$_{3}$($\theta_{3}$), as was shown in Eqs. (\ref{eq:probI}), (\ref{eq:probIyII}), and (\ref{eq:probIII}). To show the utility of the setup, we consider two cases regarding different states preparation probabilities, in which Alice can prepare the states among the cases $\eta_{1}=\eta_{2}=0.5$ and $\eta_{1}=0.3,\,\eta_{2}=0.7$ to experimentally validate Bob's discrimination device. Since in the FRIO scheme these probabilities also bias the discriminated (detected) states [see Eq. (\ref{ecu:Probb})], Alice sets these probabilities experimentally by changing the integration time of the coincidence detection rate in terms of the settled values of $\eta$ by the source. Moreover, as we mentioned before, we considered three different rates of inconclusive outcomes, namely $Q\in\{0,Q_{0}/2,Q_{max}\}$, where the specific values depend on the inner product $s$ and parameters $\eta_i$. We recall that with these three values of $Q$ correspond to MED ($Q=0$), UD ($Q=Q_{max}$) and an intermediate case ($Q=Q_)/2$) in FRIO state discrimination.

In a preliminary step, with waveplates $HWP_i(\theta_i)$ set to $\theta_i=0$, we evaluated the polarization interference visibility $\epsilon$ at the output of the interferometer by measuring in both Pauli bases $\sigma_{z}$ and $\sigma_{x}$, corresponding to the logical and diagonal polarization bases, respectively. We obtain a mean visibility $\epsilon=0.981\pm0.006$, which is typical in polarization-based experiments, where imperfections arise from experimental errors due to laser pump fluctuations in the SPDC process, imperfect spatial mode overlap and misalignment in the waveplate settings. To take these into  account in our comparison between experiment and theory, a white noise model is assumed, considering then the following state received by Bob
\begin{equation}
 \rho_{i}= \epsilon\ket{\phi_{i}}\bra{\phi_{i}} + \frac{(1-\epsilon)}{2}\mathds{1},
 \label{rho}
\end{equation}
where $i\in\{1,2\}$, $\mathds{1}$ is the identity matrix.

FRIO quantum state discrimination was implemented for state overlap $s$ ranging from orthogonal ($s=0$) to perfect overlap ($s=1$). As shown in Eq. (\ref{qubits}), the inner product $s$ is set experimentally through the angle of HWP($\theta$), since $s=\cos 4\theta$. For instance, for $s=0$ we set $\theta=\pi/8$ to generate $|\phi_1\rangle$, and $\theta=-\pi/8$ to generate $|\phi_2\rangle$. For $s=1$ we set $\theta=0$ for both $|\phi_1\rangle$ and $|\phi_2\rangle$. The experimental results are shown in Fig. \ref{iqual} (for the $\eta_1=\eta_2=0.5$ case), and Fig. \ref{diffe} (for the $\eta_1=0.3$, $\eta_2=0.7$ case). Alice used a data integration time of $10$ s for both non-orthogonal states in Fig. \ref{iqual}, while for data in Fig. \ref{diffe} times of  $6$ s and $14$ s were used for $\ket{\phi_{1}}$ and $\ket{\phi_{2}}$, respectively. In both figures we plot the experimental average success $P_{s}$, error $P_{e}$, and inconclusive probability $Q$ obtained from the recorded success $p_i$, error $r_i$, and inconclusive outcome probability $q_i$ related to the state $|\phi_i\rangle$, as shown in Eq. (\ref{ecu:Probb}). These probabilities are presented as a function of seven different values for the inner product $s\in{0,\ldots,1}$. 
In  Fig. \ref{iqual} and Fig. \ref{diffe}, the first (second) column of plots corresponds to the results obtained when Alice sends the state$\ket{\phi_{1}}$ ($\ket{\phi_{2}}$). The right-most column shows the average success, error and inconclusive probabilities, defined as the sum of these probabilities considering $\eta_{1}$ and $\eta_{2}$ [see Eq. (\ref{ecu:Probb})]. Additionally, the expected (theoretical) values of the probabilities are defined in Eq. (\ref{ecu:Probb}) using slightly mixed states \eqref{rho} and are depicted by solid lines in every plot. 

The error bars are smaller than the experimental points, and were obtained with Gaussian error propagation and considering the Poisson statistic of the recorded coincidence counts. We can observe a good agreement between the expected and recorded results for every case regarding different inner product $s$. Additionally, for the case when $\eta_1<\eta_2$ showed in Fig. \ref{diffe}, we plot the solid lines with three colors to indicate the three intervals in FRIO. Specifically, the blue color corresponds to the Interval I, while the green and red colors correspond to the Interval II and III, respectively. Although the optimal measurement depends on the interval where the FRIO discrimination is performed, we obtain results close to the expected ones regarding the three intervals. This fact is a signature of robustness of our platform against different experimental settings which allow a wide range of discrimination process contained in the FRIO scheme.

\subsection*{Conclusion}
We present a single experimental device capable of discriminating between non-orthogonal polarization states of single photons in the fixed rate of inconclusive outcomes state discrimination scheme, for which the well-known Minimum Error and Unambiguous state discrimination methods are limiting cases.  The device is based on a polarization-controlled Sagnac interferometer with nested waveplates, and allows for FRIO state discrimination to be implemented for a wide range of parameters, which are defined by the overlap and input statistics of the input states tested.  In our setup, a user Alice controls a photon pair source and encodes non-orthogonal polarization states into a heralded single photon. She sends the photon to Bob, who uses the interferometer to implement a POVM measurement with up to three outcomes.  We show that a wide range of state discrimination procedures can be implemented with this single device. Good agreement between theory and experimental results is obtained considering the two-path Sagnac interferometer visibility, which is about $98\%$. The versatility of the single device makes it directly applicable for quantum information tasks such as quantum communications.    

\bibliography{main}

\section*{Acknowledgements}
This work was supported by the Fondo Nacional de Desarrollo Científico y Tecnológico (FONDECYT) (Grant Nos. 3210359, 1180558, 1190901, 1200266, and 1200859), and the National Agency of Research and Development (ANID) -- Millennium Science Initiative Program -- ICN17$_-$012.

\section*{Author contributions}
S.G., E.S.G., S.P.W., and G.L. designed and performed the experiment. O.J. and A.D.  developed the method presented in the manuscript. S.G., E.S.G. O.J., A.D., S.P.W., and G.L. wrote and revised the manuscript. All authors gave their opinion and rewrote different sections of the manuscript.

\section*{Competing interests}
The authors declare no competing interests.
\end{document}